\documentclass[journal]{IEEEtran}
\ifdefined\pdfminorversion\pdfminorversion=7\fi
\usepackage{amsmath,amsfonts}
\usepackage{algorithmic}
\usepackage{algorithm}
\usepackage{array}
\usepackage[caption=false,font=normalsize,labelfont=rm,textfont=rm]{subfig}
\usepackage{textcomp}
\usepackage{stfloats}
\usepackage{url}
\usepackage{verbatim}
\usepackage{graphicx}
\usepackage{cite}
\usepackage{xcolor}

\begin{document}

\title{Toward Service-Balanced ISAC: From Coupled RAN to De-Coupled RAN}

\author{
Gangyong Zhu, \IEEEmembership{Graduate Student Member, IEEE},
Linghui Miao, \IEEEmembership{Graduate Student Member, IEEE},
Boyang Ji, \IEEEmembership{Graduate Student Member, IEEE},
Jiahui Liang, \IEEEmembership{Graduate Student Member, IEEE},
Shijian Gao, \IEEEmembership{Member, IEEE},
Weicheng Zhang, \IEEEmembership{Member, IEEE},
Xuesong Cai, \IEEEmembership{Senior Member, IEEE},
Jianan Zhang, \IEEEmembership{Member, IEEE},
Wei Wang, \IEEEmembership{Senior Member, IEEE},
Xiang Cheng, \IEEEmembership{Fellow, IEEE},
and Quan Yu, \IEEEmembership{Fellow, IEEE}
\thanks{Gangyong Zhu, Linghui Miao, Boyang Ji, Jiahui Liang, and Shijian Gao are with the Internet of Things Thrust, The Hong Kong University of Science and Technology (Guangzhou), Guangzhou 511453, China.}
\thanks{Weicheng Zhang, Xuesong Cai, Jianan Zhang, and Xiang Cheng are with the State Key Laboratory of Photonics and Communications, School of Electronics, Peking University, Beijing 100871, China.}
\thanks{Wei Wang is with the School of Information Science and Technology, Harbin Institute of Technology, Shenzhen 518055, China}
\thanks{Quan Yu is with Peng Cheng Laboratory, Shenzhen 518000, China.}
}

\maketitle

\begin{abstract}
Sixth-generation (6G) applications require radio access networks (RANs) to support reliable communication and seamless sensing across their operating regions. In coupled RAN deployments, shared downlink transmitting and uplink receiving sites constrain the network's ability to accommodate asymmetric links and different sensing geometries. De-Coupled RAN (DC-RAN) separates these functions, allowing independently deployed and coordinated base stations to extend uplink and downlink communication and sensing coverage. How this flexibility translates into balanced communication and sensing services, however, remains insufficiently explored. This article revisits the evolution from coupled to DC-RAN from the perspective of service-balanced integrated sensing and communication (ISAC). It examines how architectural choices affect the availability of both services, with communication-sensing coverage symmetry capturing their spatial alignment under application-specific quality requirements. Practical challenges include preserving communication consistency, maintaining sensing continuity, and coordinating distributed resources. Two case studies illustrate how DC-RAN can support coverage symmetry alongside consistent communication, and how complementary observations can sustain continuous and accurate sensing. These examples inform a discussion of future research toward service-balanced ISAC.
\end{abstract}


\section{Introduction}

Sixth-generation (6G) networks are expected to support applications that depend on both communication and sensing as native network services~\cite{itur2023imt2030}. Embodied robots, unmanned aerial vehicles, and mobile industrial platforms need to upload observations, receive coordination commands, and maintain environmental awareness under energy and mobility constraints~\cite{etsi2025isaccases}. Integrated sensing and communication (ISAC) supports data exchange and environmental observation through shared wireless infrastructure~\cite{meng2025topologies}. For a robot relying on network-assisted perception, reliable connectivity alone is insufficient if sensing becomes unavailable along its route. Both services must remain available throughout the operating region.

Meeting these demands exposes limitations of coupled radio access networks (RANs). 
{Although centralized RAN (C-RAN) enables centralized baseband processing and coordination across distributed radio sites, conventional deployments generally retain physical coupling between downlink transmitters and uplink receivers.}
{Such coupling limits the flexibility to adapt the two directions independently to asymmetric link conditions. A terminal may experience a strong downlink while its uplink remains weak~\cite{zhao2023uplink,yu2023downlink}.} Sensing further requires illumination, target visibility, and observation geometry, which communication-oriented site selection may not provide~\cite{han2025network}. Architectural evolution offers greater flexibility for coordinating distributed radio sites through shared baseband processing and control/data separation~\cite{zhang2024handover}, with the Third Generation Partnership Project (3GPP) technical specification (TS)~38.401 standardizing central unit (CU) and distributed unit (DU) separation and distinct control-plane and user-plane functions within the CU~\cite{3gpp2024architecture}. {These splits support distributed radio coordination. For cell-free cooperation, network-level ISAC has been investigated across antenna topologies ranging from massive to cell-free multiple-input multiple-output (MIMO), with communication rates and localization-error bounds characterized~\cite{meng2025topologies}. Base station (BS) placement has also been optimized to improve spatially averaged sensing accuracy under an average downlink-rate constraint, with sensing coverage further evaluated~\cite{meng2026planning}. But these studies do not examine the spatial alignment of qualified bidirectional communication and sensing coverage with independently deployed transmitters and receivers.}

De-Coupled RAN (DC-RAN) allows independently deployed downlink transmitting and uplink receiving BSs~\cite{yu2019fdran}. Thus, cooperative uplink reception can strengthen weak links and accommodate asymmetric traffic~\cite{zhao2023uplink,yu2023downlink}. Besides, independently placed receivers collect BS- or user-originated signals from complementary geometries, which can improve detection, localization, and tracking~\cite{han2025network,xue2026fdran,liu2026cooperative,meng2025cooperative}.

{This deployment freedom raises a service-level question: how can communication and sensing jointly serve a region with consistent user rates and continuous sensing?} We call the spatial alignment of regions meeting their respective quality requirements \emph{communication-sensing coverage symmetry}, measured by their intersection over union (IoU). Service balance additionally requires consistent communication across users and continuous, accurate sensing under changing observations. Channel acquisition, synchronization, and resource coordination determine whether these objectives can be sustained.

This article connects DC-RAN's deployment freedom with service-qualified coverage and practical coordination challenges. {Two case studies examine coverage-oriented deployment followed by user-oriented resource configuration, and complementary observations for localization.} Future directions address how distributed cooperation can meet evolving application needs.

\begin{figure*}[t]
    \centering
    \includegraphics[width=\textwidth]{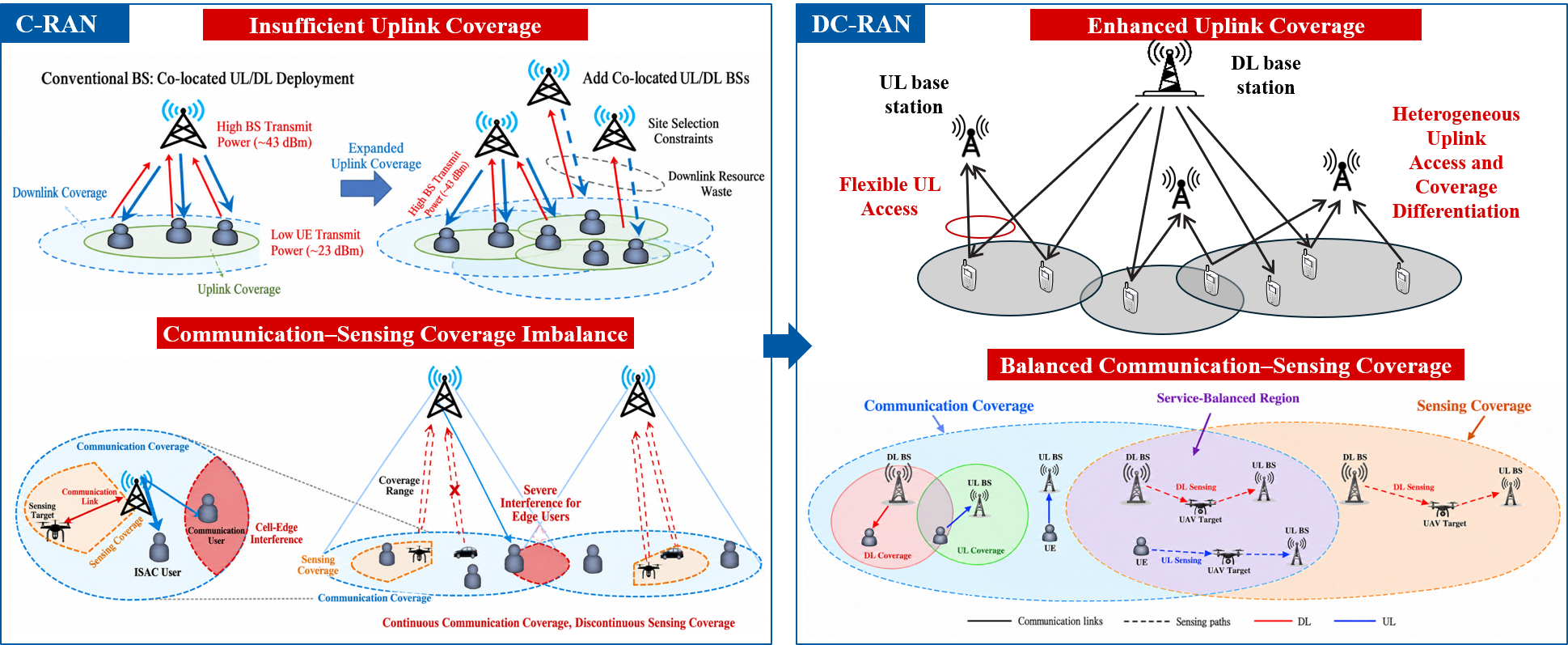}
    \caption{{Coupled radio access and DC-RAN.} {Colocated transmitters and receivers constrain directional coverage shaping. Independently deployed BSs enable flexible uplink access and complementary sensing views.}}
    \label{fig:ran-evolution}
\end{figure*}

The remainder of this article is organized as follows.
Section~\ref{sec:architecture-coverage} introduces DC-RAN and service-qualified coverage. Section~\ref{sec:shaping-challenges} explains coverage shaping and examines the coordination challenges. Section~\ref{sec:case-studies} presents the case studies, followed by future directions and conclusions in Sections~\ref{sec:future-directions} and~\ref{sec:conclusion}.

\section{Service Coverage Definition in DC-RAN}
\label{sec:architecture-coverage}

{In this section, we introduce DC-RAN and distinguish deployment from radio-resource policy. We then define communication and sensing coverage and quantify their spatial alignment.}

\subsection{DC-RAN Architecture and Network Configuration}

DC-RAN uses physically separate downlink transmitting and uplink receiving BSs with independently chosen locations and densities~\cite{yu2019fdran}. {Downlink BSs deliver data and illuminate targets, while uplink BSs receive user transmissions and echoes.} Serving and cooperation sets depend on the application's communication and sensing needs, as shown in Fig.~\ref{fig:ran-evolution}.

{A network-side service anchor associated with the CU could maintain each terminal's session across changing BS sets. A Cybertwin service agent could represent its traffic and quality-of-service (QoS) requirements. Multipath QUIC (MP-QUIC) provides a transport-design reference for managing paths between a terminal and this anchor. Mapping these paths onto separate uplink/downlink BS sets requires cross-layer coordination. Such anchoring could connect application demands with communication scheduling and required sensing updates.}

{For a service region $\mathcal{A}$, $\mathbf{z}=(\mathcal{B}_{\mathrm{DL}},\mathcal{B}_{\mathrm{UL}})$ specifies BS positions and antenna configurations. The configuration $\Phi=(\mathbf{z},\boldsymbol{\pi})$ includes radio-resource policy $\boldsymbol{\pi}$. Coverage follows prescribed rules $\boldsymbol{\pi}$ for association, cooperation, beams, power, and resources, under fixed propagation, traffic, and observation assumptions.}

\subsection{Communication Coverage}

{Bidirectional communication covers candidate locations meeting both directional requirements under the prescribed policy.} Let {$\mathcal{C}_{\mathrm{UL}}(\mathbf{z},r_{\mathrm{UL}})$} and {$\mathcal{C}_{\mathrm{DL}}(\mathbf{z},r_{\mathrm{DL}})$} denote regions meeting uplink and downlink rate thresholds. With $\mathbf{r}=(r_{\mathrm{UL}},r_{\mathrm{DL}})$,
\begin{equation}
\mathcal{C}(\mathbf{z},\mathbf{r})
=\mathcal{C}_{\mathrm{UL}}(\mathbf{z},r_{\mathrm{UL}})
\cap\mathcal{C}_{\mathrm{DL}}(\mathbf{z},r_{\mathrm{DL}}).
\label{eq:communication-coverage}
\end{equation}
{Both directional thresholds qualify candidate locations under the prescribed coverage policy. They do not impose minimum rates on all simultaneously scheduled users. Multiuser quality is evaluated during resource configuration.} The rate thresholds may differ for observation uploads and control downloads.

\subsection{Sensing Coverage}

Sensing coverage contains target locations meeting a detection or localization requirement for a fixed target class, task, and propagation model~\cite{zhang2024channel}. {Detection may require a minimum detection probability at a fixed false-alarm probability. Localization may instead impose a maximum position error.} Let {$\mathcal{S}_{\mathrm{UL}}(\mathbf{z},q_{\mathrm{UL}})$} and {$\mathcal{S}_{\mathrm{DL}}(\mathbf{z},q_{\mathrm{DL}})$} denote regions qualifying through user- and BS-originated illumination, respectively, including cooperation within each mode. With $\mathbf{q}=(q_{\mathrm{UL}},q_{\mathrm{DL}})$, when either mode independently satisfies the task,
\begin{equation}
\mathcal{S}(\mathbf{z},\mathbf{q})
=\mathcal{S}_{\mathrm{UL}}(\mathbf{z},q_{\mathrm{UL}})
\cup\mathcal{S}_{\mathrm{DL}}(\mathbf{z},q_{\mathrm{DL}}).
\label{eq:sensing-coverage}
\end{equation}
A signal-to-noise ratio (SNR) threshold can represent quality when its task relationship is established.

\subsection{Coverage Symmetry}

{Communication--sensing coverage symmetry measures spatial alignment under the prescribed coverage-evaluation policy:}
\begin{equation}
\mathrm{IoU}(\mathbf{z},\mathbf{r},\mathbf{q})
=\frac{|\mathcal{C}(\mathbf{z},\mathbf{r})\cap\mathcal{S}(\mathbf{z},\mathbf{q})|}
{|\mathcal{C}(\mathbf{z},\mathbf{r})\cup\mathcal{S}(\mathbf{z},\mathbf{q})|}.
\label{eq:coverage-symmetry}
\end{equation}

IoU compares candidate user and target locations in the same domain, which is undefined when neither service has qualified coverage. Fig.~\ref{fig:ran-evolution} illustrates the intersection in purple and the union across both service regions. Identical small regions also yield unity, so IoU must be reported with the extent of joint coverage and the service thresholds.

\begin{figure*}[t]
    \centering
    \includegraphics[width=\textwidth]{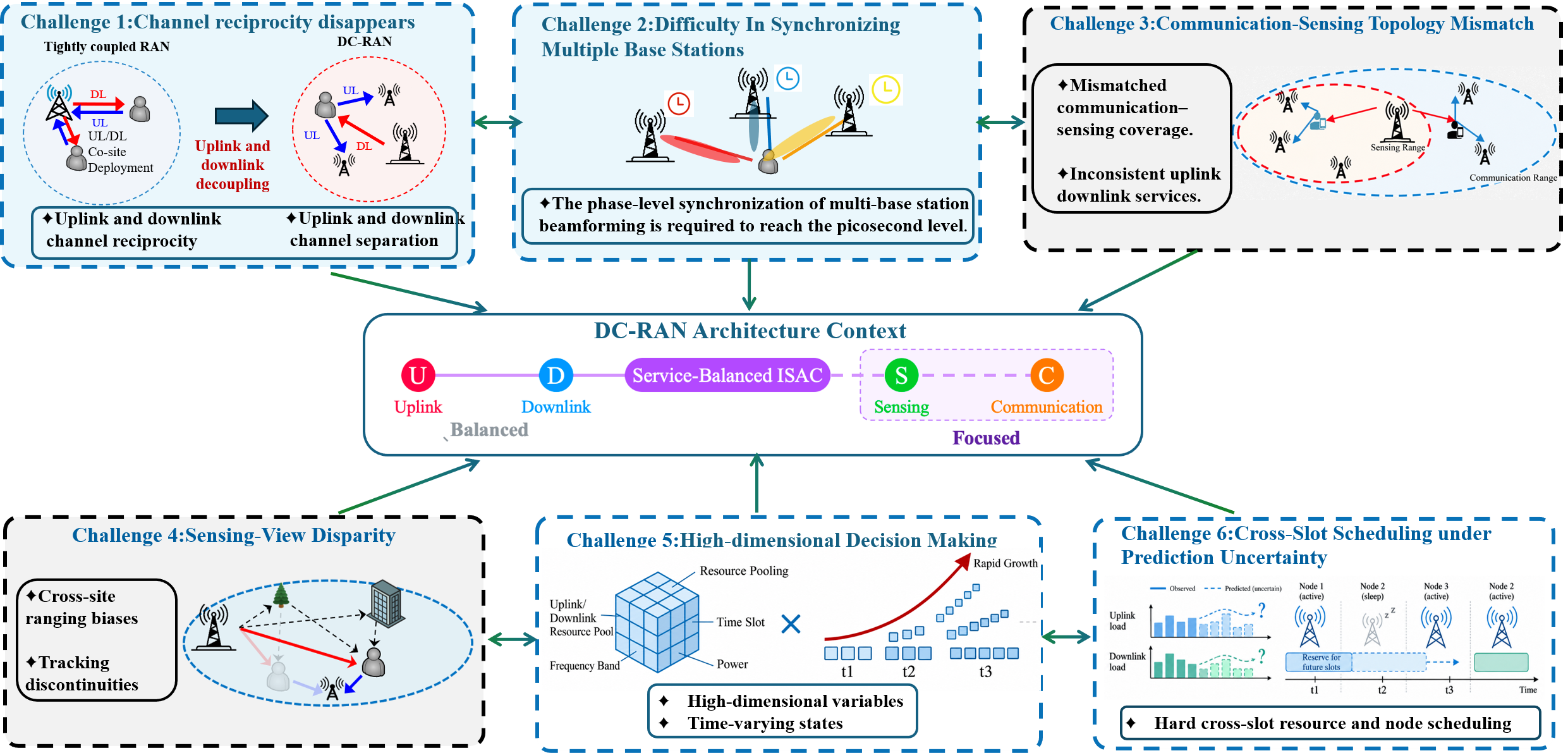}
    \caption{{Six coordination challenges in DC-RAN. Top row, left to right: channel acquisition, synchronization, and topology mismatch. Bottom row, left to right: sensing-view disparity, incomplete-state single-slot scheduling, and cross-slot decisions.}}
    \label{fig:key-challenges}
\end{figure*}

\begin{figure*}[t]
    \centering
    \includegraphics[width=\textwidth]{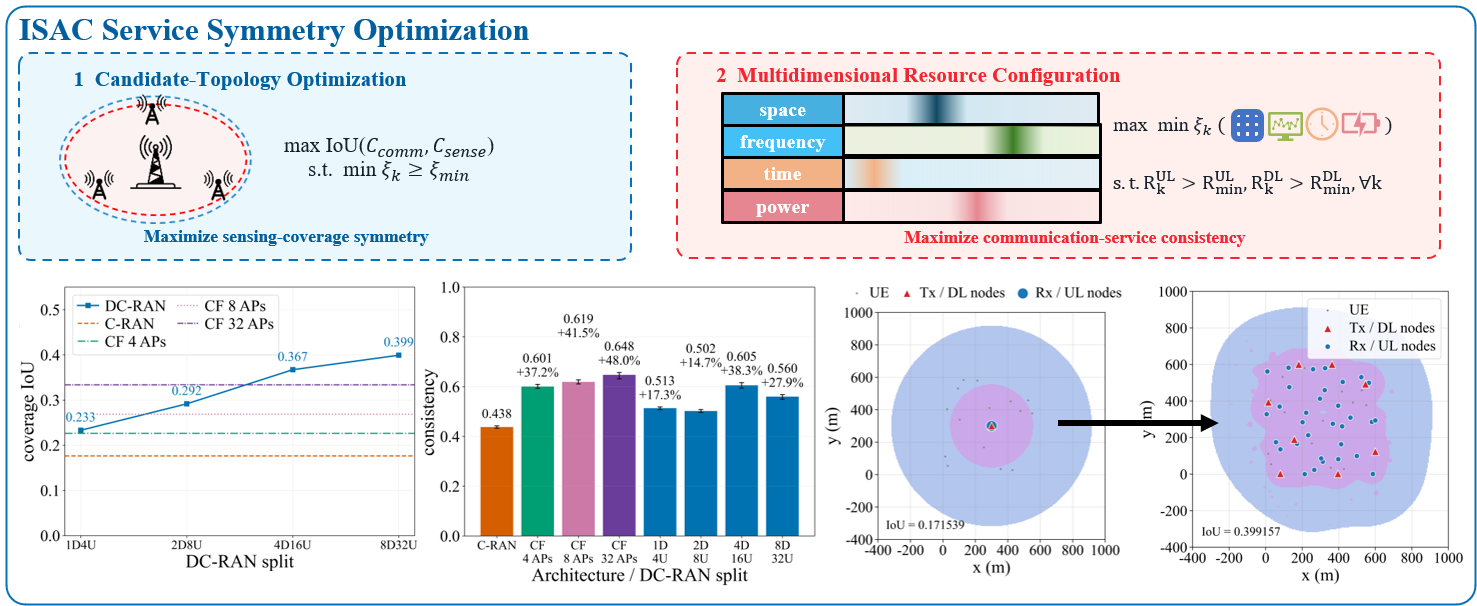}
    \caption{{Case study 1: two-stage deployment and resource optimization. Top: deployment selection and resource configuration. Bottom: communication-sensing coverage IoU, communication consistency, and spatial coverage examples.}}
    \label{fig:preliminary-results}
\end{figure*}

\section{Realizing Service-Balanced ISAC in DC-RAN}
\label{sec:shaping-challenges}

{In this section, we explain how decoupling reshapes coverage and examine the challenges of sustaining service balance: communication consistency, sensing continuity, channel acquisition, synchronization, {legacy duplexing,} and resource coordination within and across slots, as shown in Fig.~\ref{fig:key-challenges}.}

\subsection{Reshaping Communication and Sensing Coverage Through Decoupling}

Independent deployment can address the limiting communication direction: receiving BSs may be placed near power-limited users while downlink sites follow transmission needs. Cooperative reception and coordinated downlink serving sets, beams, and power can extend bidirectional coverage when both thresholds hold under a feasible allocation~\cite{zhao2023uplink,yu2023downlink}. For sensing, receiver placement changes return-path loss and observation geometry. Nearby receivers shorten echo paths, and complementary views can compensate for blockage~\cite{meng2025cooperative}. BS- and user-originated illumination provide further choices~\cite{han2025network}. A communication-served location may still lack a suitable sensing view. {Receiver or illuminator selection can then improve observation geometry. Where sensing already qualifies, changing communication serving sets or resource allocation may instead close the remaining service gap.}

\subsection{Mismatch Between Coverage Symmetry and Communication Service Consistency}

{Independent receiver placement creates a specific tradeoff: a site chosen to align sensing and communication coverage may differ from one chosen to equalize user rates.} A receiver with good target visibility may contribute little to a weak uplink user. Conversely, improving communication rates need not extend sensing throughout the communication region. Topology and allocation are interdependent: the usefulness of selected nodes depends on their feasible resources, and achievable service quality depends on both. The challenge is to improve coverage symmetry while preserving consistent rates across users and usable connectivity.

\subsection{Sensing Discontinuity Across Changing Views}

Spatial alignment does not ensure continuous sensing. {Independently placed receivers pair with BS and mobile-user illuminators, creating sensing paths with different geometries, quality, and calibration references.} Mobility and blockage change which observations are useful. {Switching views may introduce position or velocity discontinuities even when spatial coverage persists. Uncorrected timing offsets and view-dependent biases also degrade fusion.} A transition between receiving sets may preserve target visibility yet change the uncertainty of its estimated position or velocity. This makes the reliability of an observation as important as its availability.

\subsection{Limited Channel Knowledge Across De-Coupled Links}

{Separating the receiving and transmitting sites breaks the direct transfer of reciprocal-link estimates between those functions.} Uplink measurements at a receiving BS cannot directly provide downlink channel state information (CSI) for a geographically separate transmitter through reciprocity. Uplink BSs can still estimate incoming channels from pilots. In feedback-limited operation, transporting downlink measurements introduces signaling overhead and delay. Serving sets and transmission parameters chosen using incomplete or outdated information may therefore fail to deliver their expected service quality.

\subsection{Imperfect Synchronization in Multi-BS Cooperation}

Coherent joint transmission requires relative phase alignment, while uncompensated timing offsets can bias sensing and fusion. {Independently changing transmitter and receiver sets require separate reference management and bistatic timing calibration, adding coordination overhead.} Noncoherent schemes relax inter-BS phase alignment but still require appropriate symbol timing and receiver-side channel estimation. Cooperation sets and transmission modes must balance spatial diversity against achievable synchronization accuracy and its overhead.

\subsection{Computational Complexity and State Uncertainty in Single-Slot Scheduling}
{Within one slot, separate uplink/downlink associations and illuminator/receiver selections interact with frequency, transmission-time, and power allocation.} Serving choices determine available resources, while allocation determines the quality those choices achieve. These coupled discrete and continuous decisions increase scheduling complexity. Incomplete CSI further makes candidate performance uncertain: satisfying resource budgets alone does not guarantee service requirements. Scheduling must find a robust, feasible allocation within the decision deadline.

\subsection{{Cross-Slot Scheduling under Prediction Uncertainty and Duplexing Constraints}}

{Separate uplink/downlink load patterns couple current allocations to future queues, while receiver activation changes sensing visibility. Reservations and BS sleep decisions therefore affect both later communication capacity and observation availability. Mobility, traffic arrivals, and changing loads make these decisions uncertain.} {Meanwhile, legacy frequency-division duplex (FDD) and time-division duplex (TDD) procedures constrain time and frequency use across separate transmitting and receiving sites. Access, grants, feedback, and timing must remain coordinated across those sites. Logical bidirectional service can combine one-way radio legs, but its realization remains subject to spectrum, interference, and terminal capabilities. The challenge is to adapt cross-slot reservations and node activation to uncertain demand while respecting duplexing and access constraints, preserving bidirectional connectivity and sensing continuity as deployments evolve.}

\section{Case Studies of Service-Balanced ISAC}
\label{sec:case-studies}

{In this section, we examine spatial and temporal service balance through two case studies. The first evaluates coverage IoU and communication consistency. The second studies continuous localization through complementary observations.}

\begin{table}[t]
    \caption{Simulation Parameters for the Two Case Studies}
    \label{tab:simulation-parameters}
    \centering
    \small
    \renewcommand{\arraystretch}{1.15}
    \setlength{\tabcolsep}{4pt}
    \begin{tabular}{@{}>{\raggedright\arraybackslash}p{0.55\columnwidth}>{\raggedright\arraybackslash}p{\dimexpr0.45\columnwidth-2\tabcolsep\relax}@{}}
        \hline
        \textbf{Parameter} & \textbf{Value} \\
        \hline
        \multicolumn{2}{@{}l@{}}{\textbf{Common Parameters}} \\
        Deployment region & $L_x=L_y=600$~m \\
        Transmit antenna budget & $N_{\mathrm{T}}^{\mathrm{tot}}=32$ \\
        Receive antenna budget & $N_{\mathrm{R}}^{\mathrm{tot}}=32$ \\
        Carrier frequency & $f_c=3.5$~GHz \\
        Nominal downlink bandwidth & $B_{\mathrm{DL}}^{\mathrm{nom}}=100$~MHz \\
        Total downlink power & $P_{\mathrm{DL}}^{\mathrm{tot}}=43$~dBm \\
        Maximum user power & $P_{\mathrm{UE}}^{\max}=23$~dBm \\
        \hline
        \multicolumn{2}{@{}l@{}}{\textbf{{Case Study 1: Two-Stage Optimization}}} \\
        {Uplink bandwidth} & {$B_{\mathrm{UL}}=100$~MHz} \\
        {Service users / fading realizations} & {16 / 32} \\
        {Coverage window / grid spacing} & {$5.4\times5.4$~km / 5~m} \\
        {DL / UL coverage thresholds} & {2 / 2~bps/Hz} \\
        Sensing SNR threshold & $\gamma_{\mathrm{post}}^{\min}=10.80$~dB \\
        \hline
        \multicolumn{2}{@{}l@{}}{\textbf{Case Study 2: Signal Processing}} \\
        {Uplink bandwidth} & {$B_{\mathrm{UL}}=18.36$~MHz} \\
        {Scenarios / buildings} & {4 / 5} \\
        {Retained DL / UL frequency points} & {1638 / 306} \\
        {Frequency-sampling spacing} & {60~kHz} \\
        Subcarrier spacing & $\Delta f=30$~kHz \\
        Network downlink sensing symbols & $N_{\mathrm{sym}}^{\mathrm{DL}}=32$ \\
        Uplink sensing symbols & $N_{\mathrm{sym}}^{\mathrm{UL}}=16$ \\
        Power-control baseline & $P_0=-96$~dBm/PRB \\
        Path-loss compensation factor & $\alpha=0.8$ \\
        Update interval & $\Delta t=0.2$~s \\
        Trajectory duration & $T_{\mathrm{obs}}=30$~s \\
        MUSIC angular grid spacing & $\Delta\theta=0.25^\circ$ \\
        Target radar cross section & $\sigma_{\mathrm{RCS}}=10$~m$^2$ \\
        \hline
    \end{tabular}
\end{table}

\subsection{{Coverage-Oriented Deployment and User-Oriented Resource Configuration}}

{Case Study 1 first optimizes coverage-oriented deployment, then configures multiuser resources at the selected BS positions $\mathbf{z}^{\star}$.} {The simulation parameters for both case studies are listed in Table~\ref{tab:simulation-parameters}. Antenna and power budgets are matched across single-site C-RAN, distributed cell-free with 4,8,32 access points (APs), and DC-RAN deployments. Cell-free retains coupled transmitting and receiving sites~\cite{meng2025topologies}.}

{Coverage is sampled over the observation window using one noise-limited virtual terminal, maximum-ratio transmission (MRT), maximum-ratio combining (MRC), and up to four nearest BSs per direction. Sensing uses a two-leg propagation model and combined uplink/downlink coverage. Deployment selection maximizes IoU while retaining at least 90\% of C-RAN's communication coverage area. The selected deployment then remains fixed. Downlink allocation uses weighted minimum mean-square error (WMMSE) optimization. Uplink allocation uses fair power control and MMSE combining. Resource configuration retains at least 90\% of the deployment's raw sum throughput in each direction. Evaluation uses common fading realizations.}

Communication consistency is the ratio of mean rates of the weakest to strongest user quartiles, averaged equally over uplink and downlink. {Fig.~\ref{fig:preliminary-results} compares the deployments. Here, $x$D$y$U denotes $x$ transmitting and $y$ receiving BSs. IoU rises from 0.233 to 0.399 across the four DC-RAN splits. Across the evaluated deployments, higher IoU is achieved without a marked decline in communication consistency.}

\begin{figure*}[t]
    \centering
    \includegraphics[width=\textwidth]{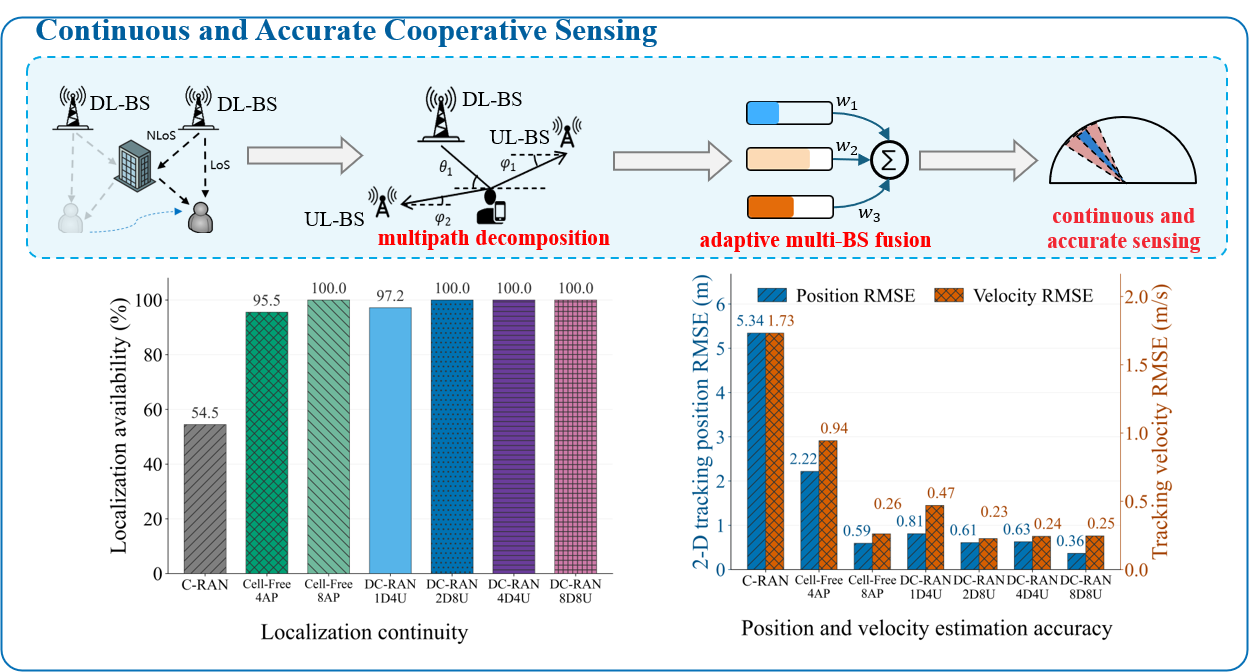}
    \caption{Case study 2. Top: multipath decomposition and adaptive multi-BS fusion. Bottom: localization continuity and position/velocity RMSE from joint uplink/downlink sensing.}
    \label{fig:continuous-localization}
\end{figure*}

\subsection{Supporting Continuous and Accurate Localization}

Fig.~\ref{fig:continuous-localization} combines multipath decomposition, uncertainty-weighted multi-BS fusion, and Kalman tracking. Receiving BSs extract delay, Doppler, and angle of arrival from direct uplink signals and vehicle-reflected downlink signals. Observations are combined according to their uncertainties, allowing the contributing links to change as mobility and blockage alter the available views. Tracking then maintains estimates between measurement updates. {Evaluation uses randomized scenarios with blockage.}

{Transmit and receive antenna capabilities are counted separately. Coupled nodes share physical arrays, whereas DC-RAN uses separate arrays.} {Sites time-share known downlink symbols, each active site using the full occupied band and shared power budget.} Uniform circular arrays use half-wavelength chord spacing and simultaneous sampling. {Sparse frequency sampling reduces processing load.} {Multiple signal classification (MUSIC) with local interpolation estimates azimuth. Uplink power follows fractional control.}

Localization availability, the fraction of snapshots producing a valid two-dimensional position update, rises from 54.5\% for C-RAN to 95.5\% for cell-free 4AP and 100\% for DC-RAN 4D4U. Root-mean-square error (RMSE) covers tracking after initial localization, including interruptions. DC-RAN 4D4U reduces position and velocity RMSE by over 70\% relative to cell-free 4AP. {Both cell-free 8AP and DC-RAN 8D8U achieve 100\% availability. DC-RAN reduces position RMSE by approximately 39\%, with a smaller velocity improvement.}

\section{Future Research Directions}
\label{sec:future-directions}

{In this section, we outline research directions for channel inference, cooperative transmission, and resource decisions under uncertainty. We also consider low-latency architectures and {incremental migration from existing deployments.}}

\subsection{Channel Acquisition and Inference via Multi-Modal Sensing}
Multi-modal sensing could assist channel acquisition through the synesthesia of machines (SoM) paradigm~\cite{cheng2024som}. Pretrained foundation models could fuse images, point clouds, user locations, and environmental maps into radio maps of link-specific path loss and multipath parameters. {Downlink inference could reduce explicit feedback, while uplink predictions could guide receiver selection and terminal power control.} {Receiver-side measurements could inform maps of geographically separate transmitter links, although direct channel reciprocity does not hold across these links.} Future research could explore which features support cooperation and how to update them under mobility, assessing prediction error, latency, feedback overhead, service quality, and terminal energy.

\subsection{Noncoherent Multi-BS Cooperative Transmission}

{Noncoherent cooperation could exploit diversity while relaxing inter-BS phase alignment. Downlink coding and interleaving could distribute data across independently modulated BS links for joint terminal decoding, retaining symbol timing and receiver-side channel estimation. Uplink observations or soft information could be aggregated at different processing stages. Joint selection of cooperating BSs, aggregation stages, and terminal power should balance user-rate consistency against transport and processing costs while retaining adequate bistatic sensing timing and calibration.}

\subsection{Robust Single-Slot Scheduling with Incomplete CSI}

{Foundation-model state estimation could combine historical channels, locations, queues, and sparse CSI into state estimates and confidence bounds. Robust optimization could jointly select separate uplink/downlink associations, sensing roles, admission, spectrum, time, and power. Budget and service-requirement checks would revise infeasible decisions or invoke a fallback, such as reducing admitted traffic or changing cooperating nodes. Research should reduce the complexity of these coupled decisions while supporting weak users and required sensing observations within the scheduling deadline.}

\subsection{Adaptive Cross-Slot Resource Orchestration}

{An adaptive controller could combine offline pretraining with online reinforcement learning. Historical observations would support self-supervised pretraining and policy distillation to initialize the controller. Online predictions of locations and directional traffic/BS loads would guide quotas, activation/sleep, and reservations while retaining needed sensing views. Throughput, delay, energy, and service-quality feedback would update the controller. Cross-slot quotas would constrain single-slot scheduling, with violations triggering conservative reservations or fallback. Research should address prediction errors and demand shifts while preserving future communication capacity and sensing continuity.}

\subsection{Accommodation for Low-Latency and High-Reliability Services}

{Meeting strict reliability and latency requirements can require additional radio and computing resources. DC-RAN could adapt cooperating BSs, directional bandwidth, and redundancy to application deadlines and delivery requirements while preserving sensing updates. Edge processing could shorten decision paths, while selective duplication could reduce dependence on retransmissions. Resource allocation should account for correlated link failures and shared transport/computing bottlenecks. Research should quantify the spectrum, energy, and coordination costs of meeting each reliability--latency target.}

\subsection{{Incremental Migration from Existing Duplexing Architectures}}

{An FDD BS could expose its transmit and receive chains as logically independent functions at the same site while retaining paired bands. A TDD deployment could retain existing sites for downlink and add low-cost receiving BSs, initially respecting configured uplink/downlink slots. Baseband unit (BBU) software and transport interfaces would coordinate access, feedback, and scheduling across these functions. Future research could assess compatibility, upgrade cost, and sensing availability. Simultaneous same-band operation requires interference management and compatible terminal capabilities.}

\section{Conclusion}
\label{sec:conclusion}

{This article has revisited the evolution from coupled RAN to DC-RAN from the perspective of service-balanced ISAC. Independent deployment and coordination accommodate distinct communication and sensing requirements. Service balance requires aligned coverage, consistent user rates, and continuous, accurate sensing. The case studies illustrate improved coverage alignment with retained communication consistency, and higher localization availability and accuracy through complementary observations. Realizing these benefits requires addressing channel uncertainty, synchronization, and resource coordination within and across slots. Future work should integrate cooperation, scheduling, and network architectures with experimental validation for 6G applications.}

\bibliographystyle{IEEEtran}
\bibliography{references}

\end{document}